# Scalable data-driven modeling of microstructure evolution by learning local dependency and spatiotemporal translation invariance rules in phase field simulation

Zishuo Lan[1], Qionghuan Zeng[1], Weilong Ma[2], Xiangju Liang[1], Yue Li[1], Yu Chen[1], Yiming Chen[1], Xiaobing Hu[3], Junjie Li[1*], Lei Wang[1], Jing Zhang[1], Zhijun Wang[1], Jincheng Wang[1*]

[1]State Key Laboratory of Solidification Processing, Northwestern Polytechnical University, Xi'an, 710072, China.

[2] School of Materials Science and Engineering, Xi'an University of Technology, Xi'an, Shaanxi 710048, China.

[3]School of Metallurgical Engineering, National and Local Joint Engineering Research Center for Functional Materials Processing, Xi'an University of Architecture and Technology, Xi'an, 710055, China.

*Corresponding authors:
E-mail address:     jchwang@nwpu.edu.cn (Jincheng Wang)
                              lijunjie@nwpu.edu.cn (Junjie Li)

## Graphical Abstract

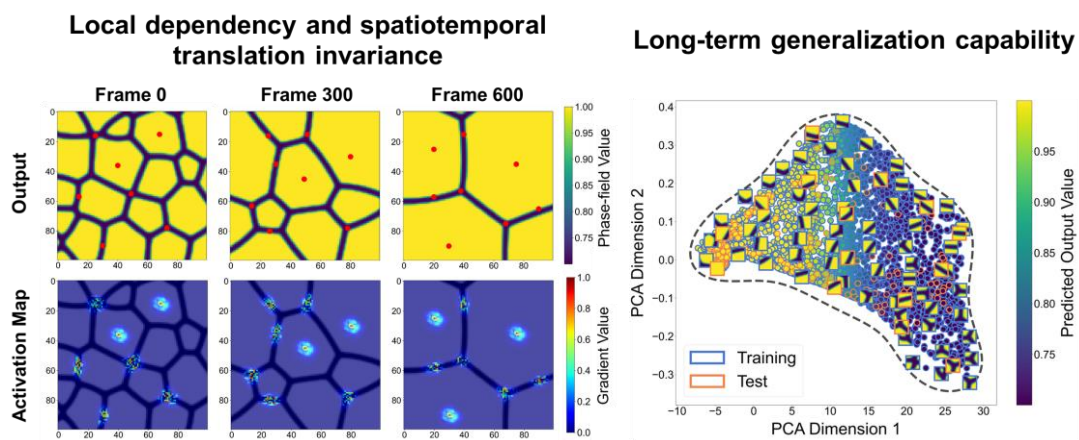

**Abstract:**

Phase-field (PF) simulation provides a powerful framework for predicting microstructural evolution but suffers from prohibitive computational costs that severely limit accessible spatiotemporal scales in practical applications. While data-driven methods have emerged as promising approaches for accelerating PF simulations, existing methods require extensive training data from numerous evolution trajectories, and their inherent black-box nature raises concerns about long-term prediction reliability. This work demonstrates, through examples of grain growth and spinodal decomposition, that a minimalist Convolutional Neural Network (CNN) trained with a remarkably small dataset—even from a single small-scale simulation—can achieve seamless scalability to larger systems and reliable long-term predictions far beyond the temporal range of the training data. The key insight of this work lies in revealing that the success of CNN-based models stems from the alignment between their inductive biases and the physical priors of phase-field simulations—specifically, locality and spatiotemporal translation invariance. Through effective receptive field analysis, we verify that the model captures these essential properties during training. Therefore, from a reductionist perspective, the surrogate model essentially establishes a spatiotemporally invariant regression mapping between a grid point's local environment (i.e., the neighboring field information that governs its evolution) and its subsequent state. Further analysis of the model's feature space demonstrates that microstructural evolution effectively represents a continuous redistribution of a finite set of local environments. When the model has already encountered nearly all possible local environments in the early-stage training data, it can reliably generalize to much longer evolution timescales, regardless of the dramatic changes in global microstructural morphology.

# 1. Introduction

Material properties are fundamentally governed by their microstructural characteristics, which makes the accurate prediction and control of microstructural evolution one of the central challenges in materials science and engineering [1–6]. Over the past three decades, the phase-field (PF) method [7–9] has emerged as one of the most powerful computational tools for predicting microstructural evolution, capable of coupling multiphysics processes and successfully capturing complex morphological transformations. This approach has provided essential modeling capabilities for understanding critical microstructural phenomena, such as solidification [10–12], grain growth [13–15], and phase precipitation[16–18]. However, the computational burden of conventional numerical solution methods, which rely on iteratively solving coupled nonlinear partial differential equations with fine spatial and temporal discretization to ensure numerical convergence, significantly limits the spatial and temporal scales achievable in simulations. Despite the application of advanced acceleration techniques, such as GPU parallelization [19,20] and adaptive mesh refinement [21,22], the computation time remains prohibitively high when simulating systems at macroscopically meaningful sizes and time scales.

In recent years, data-driven machine learning (ML) methods have demonstrated remarkable success in accelerating computationally intensive simulations [23,24]. In molecular dynamics, for instance, ML-based potential functions [25–27] have been widely adopted to extend the scale of atomistic simulations by several orders of magnitude while maintaining first-principles accuracy [28–31]. These advances have motivated researchers in the phase-field community to explore ML surrogate models [32–36]. By learning underlying patterns and evolution rules from high-fidelity PF simulation data, these models construct efficient approximation functions that can subsequently replace computationally expensive simulations for rapid predictions. Unlike traditional numerical algorithms that are constrained by strict convergence requirements, surrogate models can leap across multiple PF time steps in a single prediction step, dramatically reducing computational overhead [36–38]. This methodology has the potential to revolutionize the field by enabling surrogate models trained on small-scale, short-term PF data to rapidly predict microstructural evolution at significantly larger spatiotemporal scales [37,39].

Within the realm of data-driven acceleration for PF simulations, existing ML surrogate models are categorized into latent-space dynamics (LSD) models and pixel-space dynamics (PSD) models, differentiated by where the system's dynamics are approximated [40,41]. LSD models are based on the manifold hypothesis, which posits that microstructural evolution dynamics can be naturally approximated on lower-dimensional manifolds that exist within the high-dimensional space of raw pixel data. [35,41–44]. These models typically employ a two-step strategy: first, manifold embedding models (such as autoencoders) compress raw pixel data into a low-dimensional, information-rich latent space; subsequently, history-dependent time series models such as long short-term memory (LSTM) [45] networks learn the evolution dynamics of latent microstructural representations. For instance, Montes de Oca Zapiain et al. [35] combined autocorrelation statistics with principal component analysis to extract latent representations characterizing microstructural features, then employed LSTM to predict their evolution, demonstrating rapid prediction capabilities on spinodal decomposition microstructures. However, the reconstruction accuracy from latent representations back to microstructural images [46,47] remains limited, often producing blurred microstructures that necessitate additional short-duration PF simulations to obtain clear morphological details

In contrast to LSD models that learn evolution dynamics in low-dimensional spaces, PSD models directly approximate grid update rules of PF simulations in the pixel space [40]. This approach utilizes only a single model (such as convolutional recurrent neural networks (CRNNs) [48] or U-Net [49] architectures) that leverages multiple historical states as input to predict the evolution of each pixel in the microstructural image. For example, Yang et al. [37] applied the advanced CRNNs for video prediction (E3D-LSTM) [48] to learn the physical rules from PF data, achieving promising results in predicting grain growth, spinodal decomposition, and dendritic growth processes. In benchmark tests conducted by Dingreville et al. [40], autoencoder and LSTM-based LSD models were comprehensively evaluated against U-Net-based PSD models. Results indicated that while LSD models offer extremely fast prediction speeds suitable for screening, control, and optimization tasks with relaxed accuracy requirements, PSD models demonstrate superior prediction accuracy as surrogate models for PF simulations.

While the aforementioned models involve using multiple historical states to predict next-step microstructural evolution, PSD modeling can be further simplified to a Markovian form. This stems from the Markovian property inherent in PF simulations solved using explicit finite difference schemes, where the system state at the next timestep depends solely on the current state, independent of previous history. This property implies that ML modeling of microstructural evolution processes is actually a special case of temporal prediction, requiring consideration of only minimal historical states (i.e., the current state) for predictions. Consequently, PSD models can utilize convolutional neural networks (CNNs) [50] to predict the next state based solely on the current state, eliminating the need to introduce recurrent neural networks to model long-range temporal dependencies that do not actually exist. The effectiveness of this strategy has been confirmed by numerous studies [36,39,51–54]. For example, Alhada-Lahbabi et al. [52] applied U-Net architectures to ferroelectric PF modeling, maintaining only 5% relative error throughout complete evolution trajectories from initial random polarization states. Among these advances in PSD models, the U-Net architecture is regarded as the key to success. Its encoder-decoder architecture has naturally led researchers to interpret its performance through the lens of LSD, attributing its superior performance to the modeling of multi-scale latent manifolds that are relevant to the dynamics of microstructural evolution [38]. Guided by this concept, Bonneville et al. [54] employed U-Net as the backbone architecture and processed the latent space at the intermediate bottleneck through Vision Transformers [55] implemented in Fourier space, successfully simulating the evolution of liquid metal dealloying.

Despite these significant advances, data-driven ML surrogate models face fundamental challenges. A core principle of ML dictates that models can only make reliable predictions for samples within the training distribution. However, microstructural evolution processes described by PF equations exhibit varying evolution rates over time, with substantial morphological differences between early and late stage microstructures [37]. The LSD framework, which treats entire microstructural images as holistic units, consequently forces models to extrapolate to unseen morphological states when predicting long-term evolution [40]. Although some experimental results have confirmed that PSD models can effectively predict long-term evolution based on short-term training data [37], existing research has provided insufficient insight into the mechanisms enabling such predictions. The inherent black-box nature of these models further exacerbates skepticism regarding their prediction reliability [56]. More critically, the absence of generalization guarantees has led existing studies to collect tens to hundreds of evolution trajectories for training, even when modeling microstructural evolution under just a single set of simulation parameters [37,39,51–54]. Such requirements result in prohibitively long preparation times for high-fidelity PF data, severely limiting the practical applicability of data-driven approaches.

To address these fundamental challenges, it is crucial to recognize how the inductive biases inherent in PSD models fundamentally align with the underlying physics of PF simulations. Inductive bias refers to prior constraints embedded in ML model architectures regarding problem structure and data characteristics. Inductive biases highly compatible with modeling tasks can significantly reduce training data requirements and enable effective generalization from limited training data. Indeed, the success of nearly all modern deep learning models across various domains can be attributed to their appropriate internal inductive biases [27,50,57–60]. As a physics-based simulation method, the PF model naturally conforms to specific physical priors. Beyond the aforementioned Markovian property, PF equations describe local interactions during microstructural evolution, typically involving gradients and local free energy densities. This characteristic is clearly reflected in the finite difference solution methods employed to solve the PF equations, where each grid point's evolution is computed from its neighbors, imparting local dependency to the generated data. Simultaneously, while overall system evolution rates and morphologies change, the PF equations remain invariant throughout the entire microstructural evolution process, implying that grid update rules possess spatiotemporal translation invariance. The inductive biases of PSD models precisely align with these physical priors: CNNs implement locality and spatial translation equivariance through local connectivity and parameter sharing across spatial domains, while autoregressive inference extends equivariance to the temporal dimension by iteratively applying the same model. Moreover, unlike LSD models that treat entire images or evolution trajectories as single data units, PSD models should consider each grid point as a data unit, as all grid points evolve according to identical evolution rules. This insight suggests that model training may not necessitate collecting data from numerous evolutionary trajectories.

Building upon this foundational understanding, this study aims to demonstrate that the data requirements for training PSD surrogate models can be dramatically reduced while concurrently demystifying their generalization capabilities. First, we choose the minimalist CNN architecture, ResNet [61], with necessary modifications, and train this model using minimal PF simulation datasets for two canonical examples of microstructural evolution phenomena: grain growth and spinodal decomposition, followed by a comprehensive evaluation of its predictive reliability. Subsequently, we utilize gradient-based effective receptive field analysis [62] to verify whether the surrogate models successfully capture the physical principles of local dependency and spatiotemporal translation invariance. Finally, by analyzing the models from a reductionist standpoint, we reveal the source of their generalization capability for predicting long-term microstructural evolution.

## 2. Methods
### 2.1 Phase-field simulations

To demonstrate the general applicability of PSD models for accelerating PF simulations across diverse microstructural evolution processes, we selected two classical cases: two-dimensional grain growth and spinodal decomposition. These cases exemplify the two fundamental types of order parameter evolution equations in PF theory: the Allen-Cahn equation for non-conserved order parameters and the Cahn-Hilliard equation for conserved order parameters.

For grain growth simulations, we employed the multi-phase-field model developed by Kim et al. [14] to describe isotropic two-dimensional grain growth in zirconium metal systems. We investigated two distinct scenarios: ideal grain growth under uniform temperature conditions and non-isothermal grain growth with spatially varying temperature distributions. Separate surrogate models were developed for each scenario. To prepare the data for ML training, the multi-field outputs were converted into single-channel grayscale images, followed by data cleaning procedures to refine the final images. For spinodal decomposition simulations, we modeled the phase separation process in the Au-Pt binary alloy system using the Cahn-Hilliard equation with a subregular solution model for the chemical free energy. The resulting concentration field data, representing local Pt atomic concentration at each grid point, were directly used as single-channel images for ML model training. All simulations were performed using explicit finite difference schemes with periodic boundary conditions to generate high-fidelity training datasets. Detailed formulations of the phase-field models, simulation parameters, and data processing procedures are provided in Supplementary Material Sections S1-S3.

**2.2 Network architecture of surrogate model**

The PSD surrogate models in this study adopt the classical ResNet [61] architecture—a pure convolutional network distinguished by its residual connections that has proven effective across diverse domains. The architecture's simplicity enables a more compelling demonstration that the success of PSD models fundamentally stems from their inductive biases. Several modifications of the original architecture were implemented to meet the requirements of PF modeling, primarily including periodic boundary conditions and bounded output constraints. First, PF data employ periodic boundary conditions, but standard CNNs designed for vision tasks do not naturally preserve this periodicity, leading to additional accuracy loss at boundaries. We addressed this by replacing the default zero-padding operations before all non-1×1 convolutional layers with periodic padding (CircularPad). Second, when trained models perform long-term rollout inference, gradual error accumulation often causes prediction divergence. Previous studies employed noise adversarial training [39,52,53] to maintain microstructural morphology stability, but introducing irreducible noise can limit model accuracy in approximating deterministic PF numerical solvers. Given the natural boundedness of PF data, we added a sigmoid activation function to the final output layer and scaled it to the corresponding numerical ranges, ensuring long-term inference stability.

As shown in Fig. 1, the network accepts the current microstructural state as input to predict evolution after a fixed time interval $\Delta t_{ML}$. For ideal grain growth and spinodal decomposition, both inputs and outputs are single-channel phase fields or concentration fields. For grain growth under non-uniform temperature conditions, we compute the spatially varying normalized grain boundary mobility $\widetilde{m}(T)$ according to Eq. (S6) in the Supplementary Material and concatenate it with phase field data as model input. This preprocessing is advantageous because $\widetilde{m}(T)$ ranges from 0 to 1, which facilitates the optimization process compared to directly using temperature fields as input.

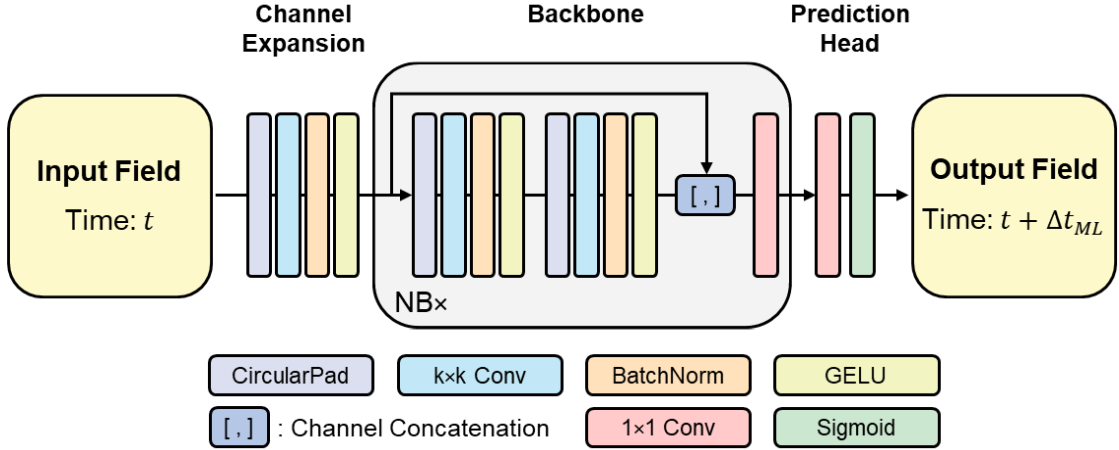

**Fig. 1.** Network architecture of the surrogate model for PF simulations acceleration.

The entire network architecture follows a modular design principle, as illustrated in Fig. 1. Unlike the U-Net, the ResNet model employed in this study does not perform downsampling, and through the implementation of CircularPad, all feature maps in the network maintain spatial dimensions consistent with the input image size. The architecture is composed of three main parts. The first part, channel expansion, passes the model inputs through a sequence of CircularPad, k×k convolution layer (k×k Conv), batch normalization (BatchNorm), and GELU activation function to elevate the channel dimension to a specified value $C$. The resulting feature maps then flow into the backbone for feature extraction, which consists of $NB$ stacked residual blocks. Each block contains two sets of CircularPad-k×k Conv-BatchNorm-GELU activation functions. To enhance gradient flow during optimization, we employ channel-wise concatenation in place of standard element-wise addition for residual connections within the blocks, followed by a 1×1 Conv to restore the original channel dimension $C$. Finally, the prediction head uses a 1×1 Conv to produce the single-channel output matching the input spatial dimensions, followed by the previously described sigmoid activation to ensure bounded predictions. Section S4 of the Supplementary Material details the architectural hyperparameters for each surrogate model trained in this work.

It is worth noting that CNNs inherently provide spatial scalability through their architecture. Since convolution operations use fixed-size kernels that can slide across spatial domains of any size, the trained model can directly process images much larger than those seen during training [37]. Combined with our periodic padding implementation, CNNs trained on small-scale microstructural evolution data can be seamlessly scaled to predict much larger systems without performance loss.

**2.3 Training strategy**

Unlike previous approaches requiring tens to hundreds of PF simulation trajectories for training, we demonstrate that PSD models can be effectively trained using merely a few small-scale simulations, revealing data requirements substantially lower than conventionally assumed. Detailed training configurations are presented in the Results section, and additional specifications are provided in Section S5 of the Supplementary Material. This study employed a one-step prediction training strategy, where models learn to predict evolution after a fixed time interval $\Delta t_{ML} = k\Delta t$, with $\Delta t$ denoting the numerical solver time step of PF simulations. Let $\delta t = h\Delta t$ represent the PF data saving interval for a simulation of total duration $t_{total} = n\Delta t$. Traditional approaches set $h = k$, where the saving interval equals the training step size. Data collected at times $t_{ik} = ik\Delta t, i \in [0, n/k]$ from a single simulation can construct training sample pairs $(t_0, t_k), (t_k, t_{2k}), \cdots, (t_{n-k}, t_n)$, yielding $n/k$ samples. To improve data utilization efficiency, we propose the overlapping training method, setting $h < k$ (where $k$ is an integer multiple of $h$) to save data at smaller intervals. Data collected at times $t_{jh} = jh\Delta t, j \in [0, n/h]$ can construct training sample pairs $(t_0, t_k), (t_h, t_{k+h}), (t_{2h}, t_{k+2h}), \cdots, (t_{n-k}, t_n)$, totaling $(n-k)/h + 1$ samples, representing approximately $k/h$ times more training data than traditional methods. We implemented this approach in ideal grain growth simulations with $h = 10$ and $k = 100$, significantly improving model accuracy without additional simulation computations. Additionally, considering the symmetry of PF data, we applied data augmentation during training using eight symmetry operations from the 2D point group 4mm, including rotations and reflections. This not only increased the number of training samples but, more importantly, embedded the corresponding symmetries into the trained model, ensuring its predictions are equivariant to these transformations.

The loss function was defined as mean squared error ($MSE$) for ideal grain growth and spinodal decomposition cases, and mean absolute error ($MAE$) for non-isothermal grain growth, quantifying the difference between predicted $ML(\phi_t^{true})$ and true phase field $\phi_{t+1}^{true}$:

$$MSE(ML(\phi_t^{true}), \phi_{t+1}^{true}) = \frac{1}{N_g} \sum_{i=1}^{N_g} (ML(\phi_t^{true}) - \phi_{t+1}^{true})^2 \qquad (1)$$

$$MAE(ML(\phi_t^{true}), \phi_{t+1}^{true}) = \frac{1}{N_g} \sum_{i=1}^{N_g} |ML(\phi_t^{true}) - \phi_{t+1}^{true}| \qquad (2)$$

Where $N_g$ is the number of grid points. The models were optimized using the AdamW optimizer [63] with a batch size of 4. Each model was trained for 150 epochs with an initial learning rate of $1 \times 10^{-3}$, which gradually decayed during training.

### 2.4 Model evaluation metrics and baseline

Previous studies commonly employ $MSE$ or $MAE$ metrics to assess one-step prediction performance [38,52,53,64], which reflects the model's ability to predict microstructural evolution at an arbitrary time. However, while such metrics measure absolute error, they lack a meaningful baseline for judging the effectiveness of the prediction. To address this limitation, we introduce the concept of an identity baseline. Consider the identity mapping model, which simply outputs its input, i.e., $ML_{id}(\phi_t^{true}) = \phi_t^{true}$, the $MAE$ between its predictions and ground truth $\phi_{t+1}^{true}$ is:

$$MAE(ML_{id}(\phi_t^{true}), \phi_{t+1}^{true}) = \frac{1}{N_g} \sum_{i=1}^{N_g} |\phi_{t+1}^{true} - \phi_t^{true}| \qquad (3)$$

This indicates that the identity model's prediction error corresponds to frame-to-frame average changes in the evolution process, and any effective predictive model should significantly outperform the identity model. Therefore, recording frame-to-frame average changes provides a natural baseline ($IB_{total}$). Comparison between the model's frame-by-frame one-step prediction $MAE$ and $IB_{total}$ measures the overall predictive ability. Since microstructure evolution primarily occurs at interfaces, we specifically use the average change of interface grid points as the baseline for interface region prediction ($IB_{interface}$). For bulk errors within grains or phases, we use the numerical precision of PF simulations (double precision in this study) as its baseline ($IB_{bulk}$).

Furthermore, we define the baseline-normalized error ($BNE$) metric as the ratio of model error to baseline over the entire evolution process:

$$BNE = \frac{\sum_{t=1}^{N_f} MAE(ML(\phi_t^{true}), \phi_{t+1}^{true})}{\sum_{t=1}^{N_f} MAE(\phi_t^{true}, \phi_{t+1}^{true})} \tag{4}$$

Where $N_f$ is the number of frames in the test set. This metric measures model performance relative to the identity baseline: values below 1 indicate the model outperforms the identity baseline, and values approaching 0 indicate excellent predictive capability.

When trained models are deployed for microstructure evolution simulations, iterative predictions must be performed via autoregressive rollout inference from initial conditions to specified time intervals. During the inference process, accumulation of minor prediction errors may cause interface position offsets between the inferred microstructural images $\phi_t^{infer}$ and ground truth, even possibly leading to partial blurring or microstructural divergence. In such cases, we no longer distinguish between bulk and interface errors, instead using the mean absolute percentage error ($MAPE$) relative to true phase fields as the evaluation metric:

$$MAPE(\phi_t^{infer}, \phi_t^{true}) = \frac{1}{N_g} \sum_{i=1}^{N_g} \frac{|\phi_t^{infer} - \phi_t^{true}|}{\phi_t^{true}} \times 100\% \tag{5}$$

Furthermore, comparison of statistical properties of microstructures obtained by surrogate models with those from PF simulations is more critical for assessing microstructural authenticity. For grain growth cases, we primarily analyze the evolution of average grain radius ⟨R⟩ and the steady-state distributions of relative grain size ⟨R⟩/R. For spinodal decomposition, we analyze the consistency of two-point spatial correlation functions (two-point statistics) [65,66]. The two-point statistics $S_2(\mathbf{r})$ is a crucial tool for characterizing spatial statistical properties of microstructures, quantitatively describing spatial distribution patterns at different length scales. This function is defined as:

$$S_2(\mathbf{r}) = \frac{1}{N_s} \sum_{i=1}^{N_s} I(\mathbf{x}_i) I(\mathbf{x}_i + \mathbf{r}) \tag{6}$$

Where $I(\mathbf{x})$ is the indicator function at position $\mathbf{x}$ (e.g., taking value 1 for high-concentration phase and 0 for low-concentration phase), $\mathbf{r}$ is the spatial vector distance, and $N_s$ is the total number of statistical samples. Here, radial averaging of $S_2(\mathbf{r})$ was used to evaluate the surrogate model's accuracy in preserving microstructural statistical features of spinodal decomposition.

**2.5 Gradient-based effective receptive field analysis**

Each position in a k×k convolutional layer's output feature map depends only on a k×k local region in that layer's input. Throughout the CNN, after multiple convolutional operations, a position in the final output depends on a larger region in the initial input, which is termed the theoretical receptive field (TRF) [62]. For networks with standard ResNet architecture, the TRF size ($s_{TRF}$) can be calculated using:

$$s_{TRF} = 1 + \sum_{i=1}^{L} (k_i - 1) \prod_{j=1}^{i-1} a_j \tag{7}$$

Where $L$ is the total number of non-1×1 convolutional layers, $k_i$ is the kernel size of the $i$-th layer, and $a_j$ is the stride of the $j$-th layer. It should be noted that the pixels within the TRF do not contribute equally to the output in actual networks. The input region with significant influence on the output is defined as the effective receptive field (ERF) [62], which is crucial for understanding the spatial dependencies of model predictions. Moreover, during training, CNN models adaptively shape their ERFs based on the inherent spatiotemporal correlations present in the data.

This paper employed gradient-based contribution analysis [62] to examine the ERF of trained surrogate models. This method quantifies input pixel contributions to output by calculating gradients of output with respect to input $g_{m,n}^{(i,j)}$ during backpropagation:

$$g_{m,n}^{(i,j)} = \frac{\partial y_{i,j}}{\partial x_{m,n}} \tag{8}$$

$g_{m,n}^{(i,j)}$ reflects the influence of input position $(m, n)$ on output position $(i, j)$. By computing $g_{m,n}^{(i,j)}$ of a specific grid point in the output image with respect to each grid point in the input image, we can obtain the ERF distribution for that output grid. We defined pixels with gradient contributions greater than 1% of the maximum gradient value as the ERF and counted their number, $n_{ERF}$. Then, the size of ERF (equivalent circular diameter) $s_{ERF}$ can be calculated as:

$$s_{ERF} = \sqrt{\frac{4n_{ERF}}{\pi}} \tag{9}$$

**2.6 Local environment feature extraction and visualization**

Drawing inspiration from atomic local environments in molecular dynamics, which characterize the chemical and geometric surroundings of atoms [25,26,67], we define the local environment of a target grid point as the multiphysics field information (e.g., phase, concentration, and temperature fields) contained within the model's ERF. The surrogate model's prediction process can be interpreted as encoding each point's local environment information into a high-dimensional feature vector (the channel dimension) through successive convolutional operations, with the prediction head subsequently mapping this feature vector to output values. This feature space is crucial for understanding the model's long-term generalization capability. To visualize this feature space, we randomly sampled grid points from training data, extracted their local environments and corresponding feature vectors at corresponding positions in feature maps before the final 1×1 convolutional layer in the prediction head. These extracted features were then visualized through principal component analysis (PCA) [68], a linear dimensionality reduction technique that identifies directions of maximum data variance (principal components). The key advantage of PCA lies in its fixed transformation parameters once trained, enabling direct projection of feature vectors from test data into the same reduced-dimensional space.

**3. Results**
**3.1 Ideal grain growth under uniform temperature conditions**

An ML surrogate model with a prediction timestep of 100Δt was trained using a single ideal grain growth PF simulation trajectory (256×256, 0–50,000Δt, saved every 10Δt) through the overlapping training methodology described in Section 2.3. The model was first evaluated on test sets with longer evolution times through one-step and rollout testing to assess training effectiveness and long-term inference performance. Subsequently, statistical property validation and quantitative verification against the Neumann-Mullins law for ideal grain growth were performed.

Fig. 2(a) presents the temporal variation of one-step prediction $MAE$ over 1,000 frames (each representing 100Δt) on test sets of 256×256, 512×512, and 1024×1024 systems, compared against the identity baseline (i.e., frame-to-frame average changes) to validate training effectiveness. All error curves exhibit consistent performance across the three system sizes, with smoother curves as system size increases, confirming seamless spatial scalability of our model architecture. The $IB_{total}$ gradually decreases over time, reflecting the reduced overall evolution rate as interfaces diminish during grain growth. The model's total error exhibits a parallel decreasing trend while remaining substantially below the baseline error. The $BNE$ calculated using Eq. (4) ranges from 0.052 to 0.054 across all three system sizes, demonstrating excellent predictive performance. Given that curvature-driven grain boundary migration constitutes the primary change during grain growth, we separately analyzed average prediction errors in interface and bulk regions, comparing them with $IB_{interface}$ and $IB_{bulk}$, respectively. As grain boundaries progressively straighten during evolution, $IB_{interface}$ decreases as shown by the red dashed line. The model's prediction error at interfaces remains over an order of magnitude lower than $IB_{interface}$, confirming reliable interface motion prediction. Errors in the bulk regions approach numerical precision levels due to sigmoid function constraints. Notably, for predictions beyond the training data temporal range (>500 frames), model errors remain stable, indicating effective long-term evolution prediction capability.

We subsequently evaluated the surrogate model's rollout inference performance for practical applications. As shown in Fig. 2(b), autoregressive rollout inference from frame 0 to frame 1000 yields progressively accumulated $MAPE$ of 0.5%, 1.2%, and 2% for the 256×256, 512×512, and 1024×1024 systems, respectively. Despite gradually increasing $MAPE$ compared to ground truth, Fig. 2(c) demonstrates that the final microstructural morphologies remain clear without unrealistic artifacts. The 256×256 system evolved by the surrogate model shows nearly perfect agreement with PF simulations, exhibiting only minor interface position offsets. Larger systems display higher errors due to structural differences in localized regions, stemming from the bifurcation of grain boundary connectivity during grain disappearance [37], leading to divergent evolution paths under identical initial conditions. Fig. 2(d) illustrates the rollout inference $MAPE$ at the final time when starting from different initial frames. The final frame $MAPE$ dramatically reduced as the inference starting frames increased from 0 to 200, followed by a gradual stepwise decrease after frame 200. This behavior arises from extensive grain disappearance during early evolution, making local evolution more susceptible to deviation from original trajectories.

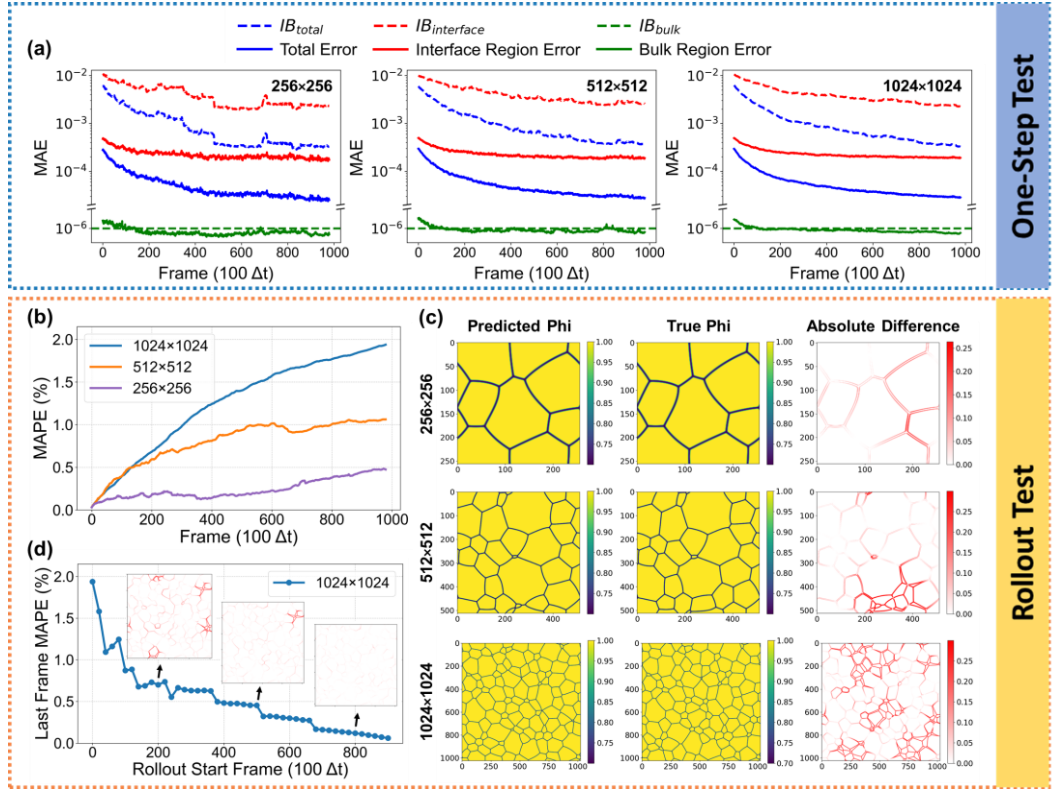

**Fig. 2.** Evaluation of the ML surrogate model for ideal grain growth. (a) Temporal variation of one-step prediction *MAE* on three test sets with different system sizes. (b) Temporal variation of *MAPE* during autoregressive rollout inference from initial frames. (c) Comparison between predicted microstructural morphologies and ground truth PF simulations at the final time (frame 1000). (d) Last frame *MAPE* for rollout inference initiated from different starting frames on the 1024×1024 system, with several inserts of final microstructural differences.

Although the surrogate model exhibits error accumulation in pixel-wise predictions during inference due to evolution path divergence, this does not necessarily compromise the physical realism of the generated microstructures. Examining whether the surrogate model reproduces statistical features of PF microstructures is more meaningful. We therefore investigate the statistical consistency between the microstructure generated by the PF simulation and that predicted by the surrogate model, which evolves independently from Voronoi polygon initial conditions. Fig. 3(a) shows the surrogate model's long-term rollout inference over 6,000 frames (equivalent to 600,000Δt, 12 times longer than the training time range) in a 256×256 system from Voronoi polygon initial conditions. The model successfully predicts complete evolution from initial polycrystalline structures to the final stable state of four hexagonal grains with 120° triple junction angles. We further analyzed grain growth kinetics in 2048×2048 systems, including temporal evolution of average grain radius and normalized steady-state grain size distributions. To ensure robust statistical validation, PF and surrogate model simulations were initiated from different random initial conditions (different initial ⟨R⟩ and the distribution of R/⟨R⟩), providing independent samples for statistical comparison. As shown in Fig. 3(b), the mean grain radius squared ⟨R⟩$^2$ from surrogate models exhibits excellent linear correlation with time (correlation coefficient $corr = 0.9985$), in perfect agreement with the parabolic growth law. The linear regression slopes closely match PF simulation results, showing only 4.8% relative deviation. Grain growth simulated by the surrogate model achieves self-similarity after frame 170, with steady-state relative grain size distributions obtained through statistics from frames 170-190. As shown in Fig. 3(c), the result aligns remarkably well with the PF simulation. These results confirm high statistical consistency between the surrogate model and PF simulations, validating the reliability of the surrogate model in simulating microstructural evolution.

We further implemented the surrogate model in single-grain growth simulations to validate its consistency with the Neumann-Mullins law [69,70]. The Neumann-Mullins law describes the quantitative relationship between the grain area change rate and the topological parameter of grain sides:

$$\frac{dA_{ns}}{dt} = \frac{\pi}{3}m\sigma(ns - 6) \tag{10}$$

Where $A_{ns}$ represents the area of an $ns$-sided grain, $m$ denotes grain boundary mobility, and $ns$ denotes the number of grain sides. This equation indicates that grains with more than six sides grow while those with fewer shrink, with area change rates proportional to the deviation from six sides, independent of time or grain size. The Neumann-Mullins law makes no assumptions about grain shape or boundary curvature and has been proven as an exact formula for two-dimensional ideal grain growth. As shown in Fig. 3(d), the surrogate model's predictions align precisely with theoretical expectations for single-grain systems: the three-sided grain shrinks, the six-sided grain remains unchanged, and the nine-sided grain expands. Furthermore, Fig. 3(e) demonstrates that area changes over time for grains with different numbers of sides closely match Neumann-Mullins law predictions, providing quantitative validation of the surrogate model's accuracy.

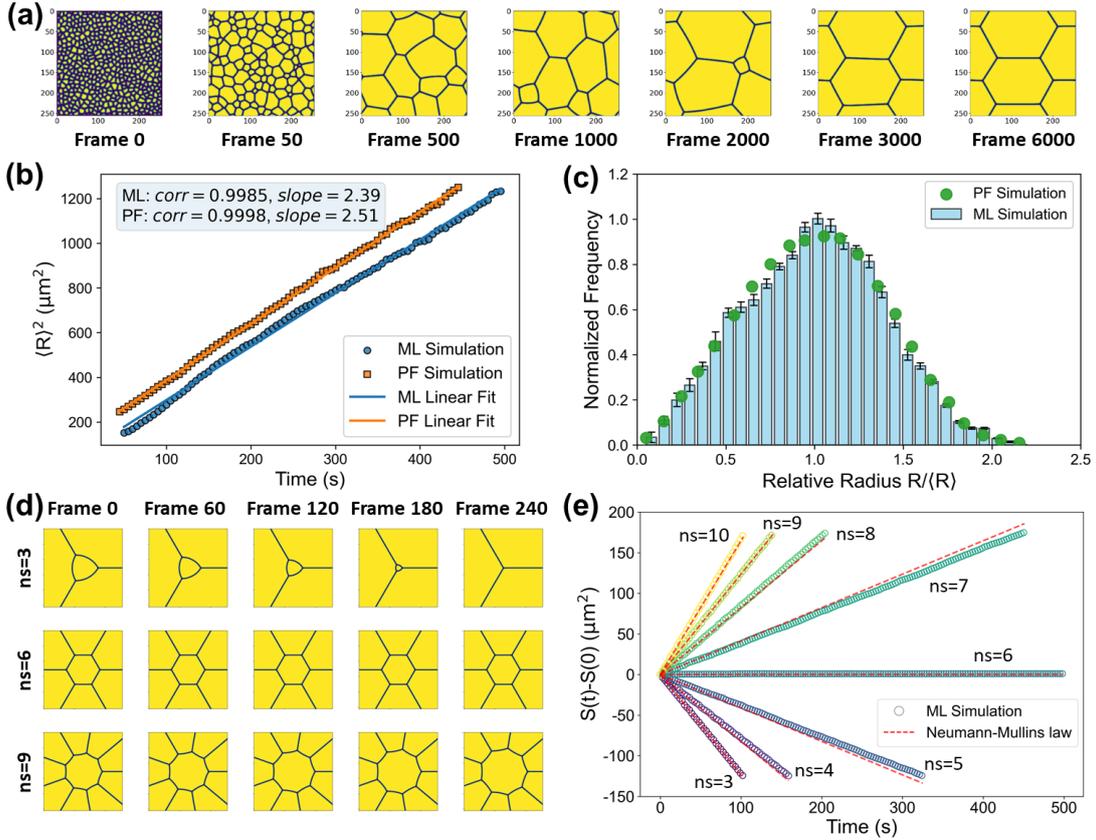

**Fig. 3.** Statistical property validation and theoretical consistency verification of the ML surrogate model for ideal grain growth. (a) Microstructural morphology evolution over 6000 frames from Voronoi initial conditions (256×256). (b) Temporal evolution of average grain radius squared in 2048×2048 polycrystalline systems,

comparing ML predictions with PF simulations from different initial conditions, including linear regression results. (c) Comparison of steady-state grain size distributions between ML predictions and PF simulations. (d) Morphological evolution of regular grains with $ns$ sides. (e) Comparison of temporal area changes for grains with varying numbers of sides against Neumann-Mullins law predictions.

### 3.2 Grain growth under non-uniform temperature distributions

Microstructural evolution in real materials often occurs under the influence of non-uniform external fields such as temperature or stress distributions. In data-driven approaches, modeling such microstructural evolution can be achieved by training surrogate models with phase field and corresponding external fields as inputs to predict subsequent phase field evolution. This section demonstrates the significant potential of ML methods for establishing surrogate models for grain growth under non-uniform temperature distributions.

We trained the surrogate model with a 100Δt timestep using PF data saved every 100Δt during 0-50,000Δt on 256×256 systems under four time-invariant temperature fields, as shown in the left part of Fig. 4(a), including vertical linear gradient (VerticalGrad), diagonal linear gradient (DiagonalGrad), single-period and double-period sinusoidal (SinePeriod1 and SinePeriod2) distributions. We tested the model over 1000 frames (each frame equals 100Δt) on both training sets and four previously unseen temperature distributions, including radial gradient (RadialGrad), Gaussian mixture (GaussianTemp), single-period and double-period checkerboard trigonometric (TrigoPeriod1 and TrigoPeriod2) distributions, as shown in the right part of Fig. 4(a). One-step prediction evaluation results (see Section S6 in Supplementary Material) indicate that the model significantly outperforms the identity baseline on most datasets. Rollout inference evaluation results are shown in Fig. 4(a). Under eight temperature field configurations from both training and test sets, final microstructural morphologies obtained through the surrogate model show excellent agreement with PF simulation results, successfully reproducing temperature field regulatory effects on microstructural evolution.

The above surrogate model trained on time-invariant temperature fields was further applied to predict grain growth under time-varying temperature conditions. We constructed a temperature field function simulating metal heating in a furnace, with an initial center temperature of 923 K and an edge temperature of 1023 K, eventually reaching a uniform 1023 K through heat conduction. The detailed settings are given in Section S2 of the Supplementary Material. Starting from Voronoi initial microstructures, we performed rollout inference over 1000 frames. Fig. 4(b) illustrates the temporal evolution of temperature fields and polycrystalline microstructures, clearly demonstrating the dynamic regulatory effects of time-varying temperature fields on grain growth. Edge regions undergo coarsening first, followed by gradual grain boundary motion in interior regions as the interior temperature increases continuously, ultimately forming gradient microstructures with coarse edge grains and fine center grains.

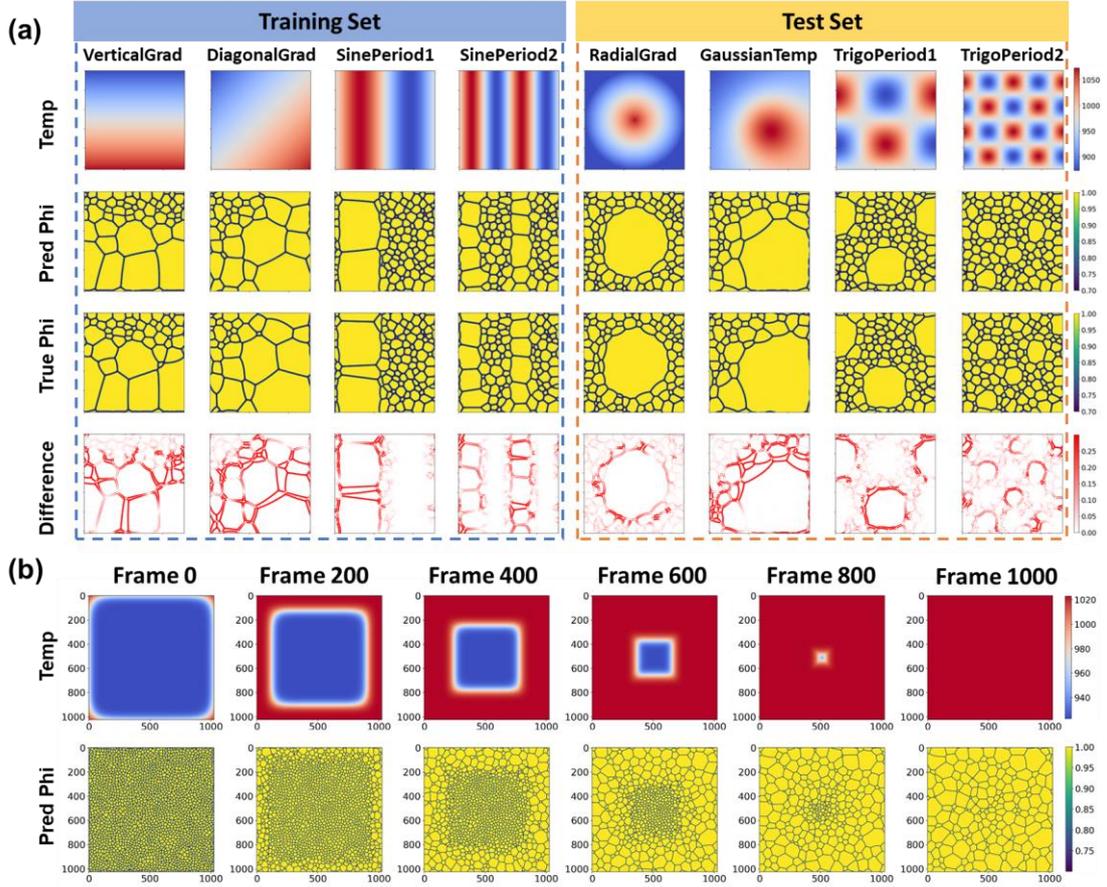

**Fig. 4.** Evaluation of the ML surrogate model for grain growth under non-uniform temperature distributions. (a) Comparison of final evolution results (frame 1000) predicted by the surrogate model with PF simulations, including temperature field configurations in both training and test sets. (b) Microstructural evolution predicted by the surrogate model from Voronoi initial conditions under time-varying temperature fields mimicking metal heating in a furnace.

### 3.3 Spinodal decomposition

Spinodal decomposition represents another common case for ML acceleration of PF simulations. The evolution process exhibits distinct two-stage characteristics: rapid phase separation in the early stage, followed by exceedingly slow coarsening. This significant timescale disparity between stages presents severe challenges for ML modeling. We first constructed a large-timestep surrogate model for the coarsening stage (after 2,000Δt). Subsequently, by analyzing this model's prediction errors, we determined appropriate timesteps for modeling the early rapid evolution stage, enabling complete microstructural evolution prediction from initial noise through final coarsening.

An ML model with a 500Δt prediction timestep was trained for the coarsening stage using a single spinodal decomposition PF simulation trajectory (256×256, 2,000-100,000Δt, saved every 100Δt) through the overlapping training methodology described in Section 2.3. Fig. 5(a) presents comparisons between one-step prediction $MAE$ and the identity baseline from 2,000Δt (frame 5) to 200,000Δt (frame 400) on test sets of 256×256, 512×512, and 1024×1024 systems. Similar to the ideal grain growth case, the error curves remain nearly identical across all three system sizes, demonstrating excellent spatial scalability. The baseline and surrogate model $MAE$ decrease rapidly during frames 5-50 before stabilizing. The measured $BNE$ values of 0.075-0.076 confirm effective predictive capability. The interface errors exhibit a rapid decrease during frames 5-50, followed by a gradual increase; however, these errors remain significantly below $IB_{interface}$, indicating effective long-term interface prediction.

Rollout test results from frame 5 to final frame 400, shown in Fig. 5(b), indicate cumulative $MAPE$ of 4.6%, 3.5%, and 4.2% for the 256×256, 512×512, and 1024×1024 systems, respectively. Microstructural morphology comparisons in Fig. 5(c) reveal that despite structural differences in limited local regions due to chaotic effects, surrogate model predictions closely match PF simulation results in most regions, with only minor interface position offsets. Fig. 5(d) shows rollout inference $MAPE$ at the final time when starting from different initial frames. Inference starting from frame 15 yields significantly reduced final errors compared to that starting from frame 5, followed by a gradual decline with later starting frames. This indicates the surrogate model's difficulty in predicting dramatic early coarsening stages. Notably, when this surrogate model begins inference from frame 100 (50,000Δt) onward, final microstructural morphologies show virtually no local structural differences from PF simulation results, exhibiting only minor interface position deviations. Therefore, we trained two additional models with smaller timesteps for the early coarsening stage (2,000-50,000Δt) and phase separation stage (0-2,000Δt).

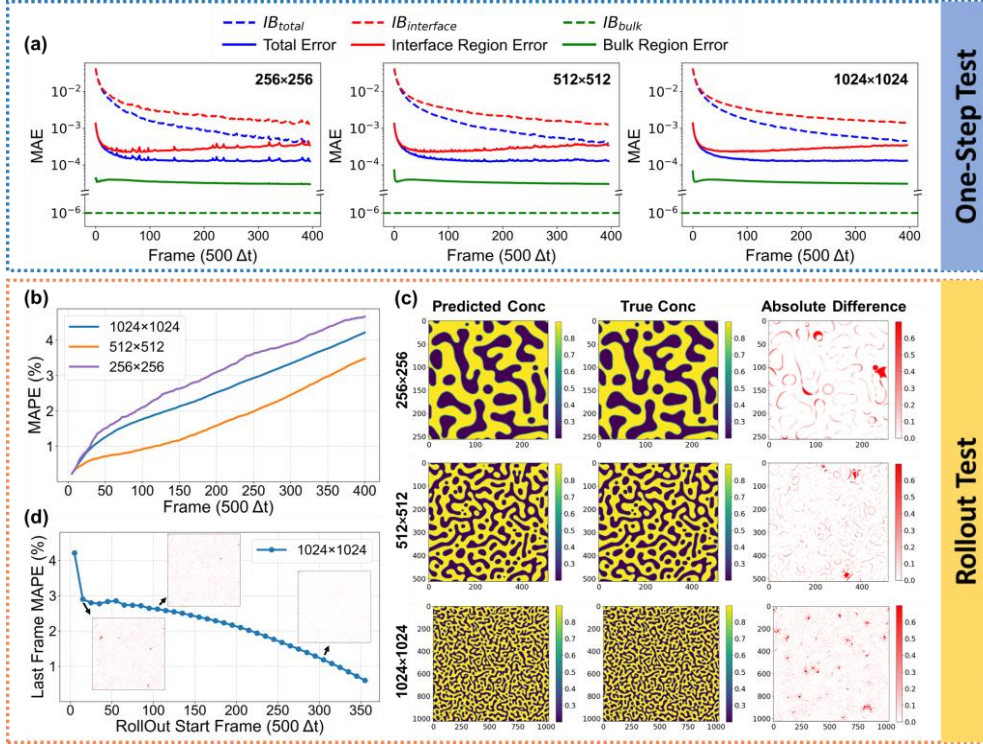

**Fig. 5.** Evaluation of the ML surrogate model for spinodal decomposition. (a) Temporal variation of one-step prediction $MAE$ on three test sets with different system sizes. (b) Temporal variation of $MAPE$ during autoregressive rollout inference from initial frames. (c) Comparison between predicted microstructural morphologies and ground truth PF simulations at the final time (frame 400). (d) Last frame $MAPE$ for rollout inference initiated from different starting frames on the 1024×1024 system, with several inserts of final microstructural differences.

While extensive studies have pursued ML acceleration of spinodal decomposition evolution, the extreme intensity of phase separation stages has prevented the development of ML models capable of evolving directly from initial noise states [35,37,42,71]. This limitation severely restricts surrogate model applications as independent simulators. To achieve complete evolution prediction from initial noise states, we trained two additional models with timesteps of 10Δt and 100Δt for phase separation and early coarsening stages, respectively. Combined with the aforementioned 500Δt model, a three-stage sequential prediction system was constructed: the 10Δt model handles phase separation from noise to 2,000Δt, the 100Δt model covers early coarsening from 2,000Δt to 50,000Δt, and the 500Δt model manages late coarsening from 50,000Δt to 200,000Δt. Detailed model hyperparameters and training configurations for each stage are provided in Sections S4 and S5 of the Supplementary Material.

Fig. 6(a) and (b) present one-step prediction evaluation results for the 10Δt and 100Δt models on 1024×1024 test sets, respectively. The 10Δt model's baseline and prediction errors exhibit distinct three-stage characteristics: decreasing during the noise relaxation stage (0–20Δt), increasing during the phase separation stage (20–600Δt), and continuously decreasing during subsequent coarsening. The 10Δt model prediction errors maintain approximately one order of magnitude lower than the baseline, with $BNE$ of 0.064 indicating effective prediction of complex phase separation evolution. Fig. 6(b) shows that the 100Δt model specialized for early coarsening prediction exhibits more gradual changes in both baseline and prediction error compared to the 500Δt model shown in Fig. 5(a). This reduces learning difficulty and enables more accurate predictions during this period of rapid microstructural changes. The results of the complete three-stage sequential rollout prediction are presented in Fig. 6(c) and (d). For the 1024×1024 system, the final cumulative $MAPE$ reaches approximately 17.5%. This larger error is expected, as initiating inference from a random noise state amplifies minor early-stage prediction inaccuracies over the long-term evolution. Nevertheless, microstructural morphology comparisons throughout the evolution process shown in Fig. 6(d) demonstrate that surrogate models successfully reproduce complete evolution from phase separation to coarsening, with final microstructural morphology highly similar to PF simulation results.

To validate the physical realism of microstructures predicted by the surrogate model, we conducted quantitative comparisons of statistical features against ground truth. Fig. 6(e) and (f) present radially averaged two-point statistics $\bar{S}_2$ comparisons between the surrogate model and the PF simulation at 50,000Δt and 200,000Δt, respectively. At 50,000Δt, the two results nearly overlap completely, indicating accurate microstructural statistical feature prediction. However, at 200,000Δt, observable differences emerge. The two radial distribution curves show highly synchronized oscillation patterns, confirming good agreement in both the average critical feature size (ACFS, identified by the first minimum) and the average feature-to-feature distance (AFFD, identified by the second maximum) [42,43]. The two curves exhibit differences primarily in vertical displacement. The two-point statistics function describes the probability of two points separated by distance $r$ belonging to the same phase. As $r \to 0$, $\bar{S}_2$ equals the phase volume fraction (PVF), and as $r \to \infty$, $\bar{S}_2$ approaches the square of PVF, indicated by horizontal dashed lines in Fig. 6(f). Thus, the two-point statistics differences in Fig. 6(f) primarily result from differences in PVF, with ground truth at 54.14% versus the surrogate model's prediction of 56.23%. Fig. 6(g) presents the temporal evolution of PVFs for both the PF simulation and surrogate model prediction. The 500Δt surrogate model gradually deviates from true PVFs during long-term inference, ultimately producing approximately 2% deviation. This phenomenon reflects an important limitation of our surrogate model: while model outputs satisfy physical boundedness constraints, they fail to naturally satisfy concentration field conservation constraints. Developing models that satisfy such conservation constraints represents an important future research direction.

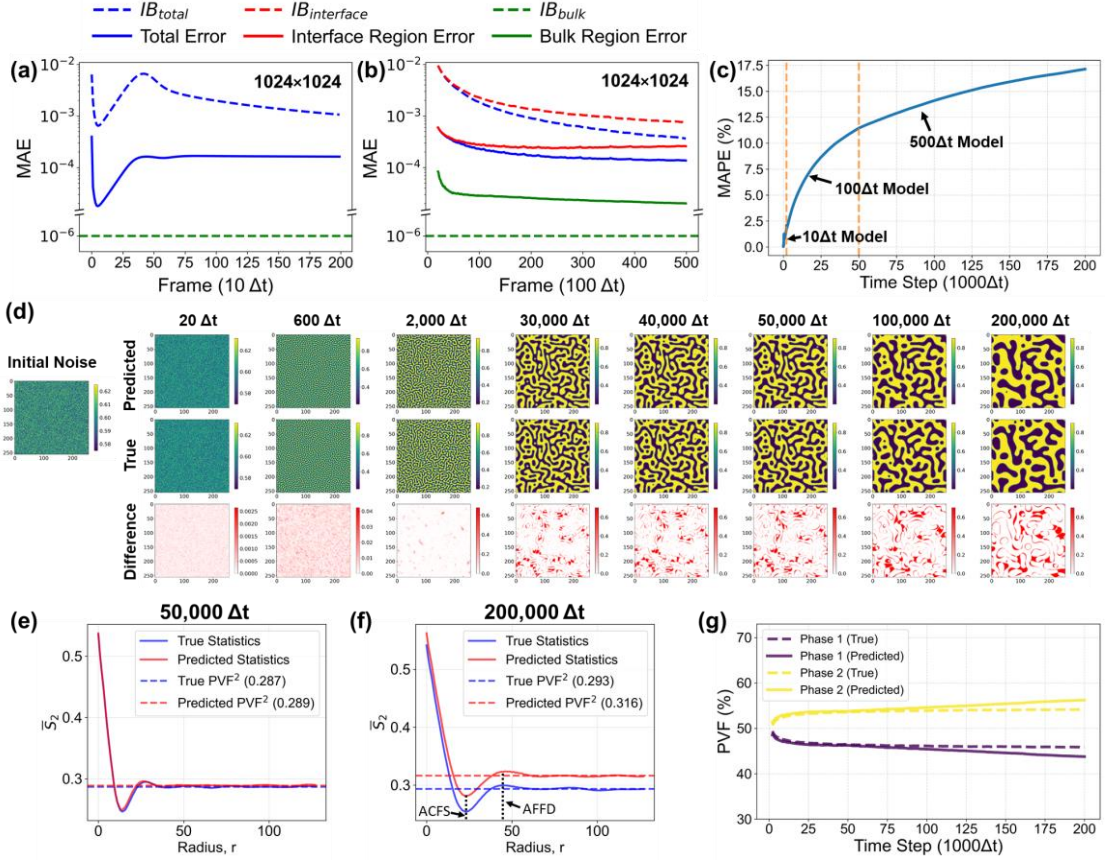

**Fig. 6.** Three-stage sequential prediction of complete spinodal decomposition evolution. (a) One-step prediction $MAE$ of the 10Δt model. (b) One-step prediction $MAE$ of the 100Δt model. (c) $MAPE$ accumulation during three-stage model autoregressive rollout inference from noise to 200,000Δt (1024×1024). (d) Comparison of microstructural morphology predicted by the three-stage surrogate model with PF simulation (256×256). (e-f) Two-point statistics comparisons at 50,000Δt and 200,000Δt. (g) Phase volume fraction evolution comparison.

## 4. Discussion
### 4.1 Local dependencies and spatiotemporal invariance captured by surrogate models

PF equations fundamentally describe microstructural evolution as a process characterized by pronounced local dependencies and spatiotemporal translational invariance. In this section, we employ gradient-based ERF analysis to demonstrate that our surrogate models have indeed learned these fundamental physical principles through their architectural inductive biases. As described in Section 2.5, for a CNN with identical input and output dimensions, each output pixel's prediction is theoretically influenced by a region in the input known as the TRF. The size of TRF ($s_{TRF}$) is determined solely by the network's hyperparameters, as detailed in Eq. (7). However, only a subset of pixels within the TRF—termed the ERF—effectively contributes to the final prediction. The contribution of each input pixel can be quantified by computing the gradient of a specific output pixel's value with respect to every input pixel, where a higher gradient magnitude signifies greater influence. The spatial distribution of these gradient magnitudes forms an activation pattern that directly visualizes the ERF.

Fig. 7(a) illustrates the gradient contribution analysis for the 100Δt ideal grain growth model ($s_{TRF} = 35$), as described in Section 3.1. We sampled multiple grid points at characteristic locations (grain interiors, grain boundaries, and multi-grain junctions) across different temporal frames, as marked by red points in the output images of Fig. 7 (a). For each sampled point, we computed normalized gradients with respect to all input pixels and overlaid the resulting activation heatmaps onto the input microstructures. The visualizations reveal highly localized activation patterns for all sampled points, confirming that model predictions rely on local information. Remarkably, sampling points within grain interiors exhibit nearly identical activation patterns regardless of their temporal or spatial positions. This consistency demonstrates that the model applies uniform computational operations at these locations, providing direct evidence that it has captured the spatiotemporal translation invariance inherent to the physical process. Activations at grain boundaries present a narrow, elongated shape that covers the local boundaries, suggesting that the model implicitly captures local interface curvature, the primary driving force for grain boundary migration. At multi-grain junctions, activation regions become broader and more dispersed, enabling the model to integrate the complex geometric information necessary for accurately predicting topological events, such as grain switching or annihilation, that occur at these sites.

We subsequently computed gradients for numerous grid points across multiple temporal frames and calculated averages for each grid point category (grain interior, grain boundary, and triple junction). The results are presented in Fig. 7(b), where the left column displays the full TRF region and the right column shows magnified views of the central ERF regions. For all grid point types, significant gradient values are confined to a small central fraction of the TRF, indicating substantially smaller ERFs relative to the TRFs. Following the method described in Section 2.5, statistical analysis of all grid points reveals that the ideal grain growth model exhibits an average ERF diameter of 12.3 pixels, which occupies merely 9.7% of the TRF area. Similar analyses for non-isothermal grain growth and spinodal decomposition models reveal the same findings, as detailed in Section S7 of the Supplementary Material. These results provide compelling evidence that surrogate model predictions depend on local neighborhood states. This finding underscores a fundamental distinction between PF surrogate modeling tasks and traditional computer vision tasks, such as image classification or object detection, where ERFs typically span much larger regions to integrate global contextual information for accurate predictions [62].

The non-isothermal grain growth surrogate model described in Section 3.2 accepts two input channels—the phase field and temperature field—which jointly determine the evolution of the phase field. Higher temperatures increase grain boundary mobility, accelerating interface migration. To investigate temperature effects on ERFs, we analyzed $s_{ERF}$ as a function of local temperature for both input channels. Results in Fig. 7(c) demonstrate that increasing temperature produces clear ERF expansion in both phase field and temperature field input channels. This phenomenon reveals two critical insights. First, at higher temperatures, interfaces migrate faster and microstructural changes propagate over larger distances within a single timestep, necessitating information from broader spatial neighborhoods for accurate prediction. Second, the model has successfully learned to dynamically adjust its information-gathering range according to local temperature conditions, automatically expanding its ERF in high-temperature regions where rapid evolution occurs. This adaptive behavior confirms that the surrogate model has captured the temperature-dependent dynamics inherent in non-isothermal grain growth.

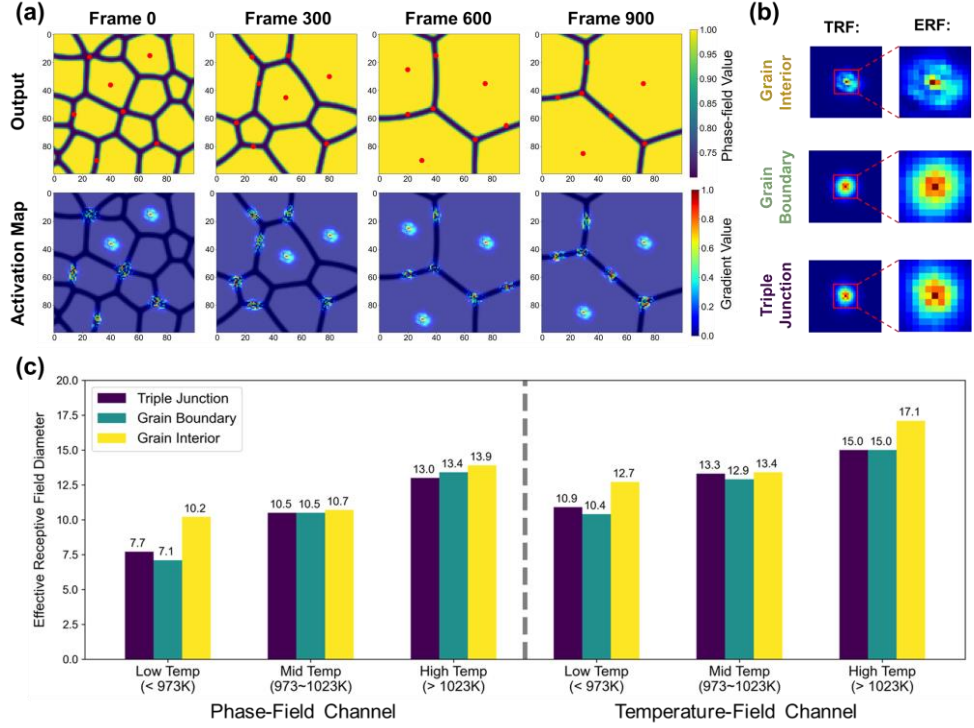

**Fig. 7.** ERF analysis of grain growth surrogate models. (a) Gradient contribution analysis for the multi-frame evolution of ideal grain growth. The top row illustrates model output microstructures with sampled points (marked in red), while the bottom row shows the corresponding gradient activation heatmaps overlaid on the input microstructures. (b) Averaged gradient contributions for different grid point types across multiple time frames. (c) Temperature-dependent variation of $s_{ERF}$ for the grain growth model under non-uniform temperature distributions.

To investigate the determinants of ERFs, we conducted systematic experiments using three groups of models with different timesteps (10Δt, 100Δt, and 1000Δt) trained on the ideal grain growth dataset. Each group employed an identical series of network architectures with varying depths—deeper networks possess larger TRFs. Detailed experimental settings are provided in Section S8 of the Supplementary Material. Fig. 8(a) presents the relationship between $s_{ERF}$ and $s_{TRF}$ for the three model groups. The results demonstrate that models trained with larger timesteps possess larger ERFs, consistent with the intuition that larger timesteps require broader neighborhood information to support predictions. Moreover, $s_{ERF}$ does not increase indefinitely with $s_{TRF}$ but gradually approaches an upper limit determined by the timestep size. This saturation behavior indicates that the spatial extent of information required for predicting each grid point's evolution is governed by the physical laws embedded in the training data rather than model hyperparameter configurations.

Fig. 8(b) shows the corresponding average one-step prediction error variations with $s_{TRF}$ for the three model groups. As $s_{TRF}$ and model parameter counts increase, prediction errors gradually decrease and substantially outperform corresponding identity baselines. When analyzed in conjunction with the $s_{ERF}$ variation in Fig. 8(a), it becomes evident that before ERFs reach their upper limits, model error reduction exhibits markedly steeper slopes, indicating that insufficiently small TRFs significantly constrain model performance. After ERFs reach saturation, error reduction trends notably decelerate, with further parameter increases yielding limited performance improvements. This implies that initial model architectures should possess sufficiently large TRFs to accommodate the modeling task requirements for optimal performance.

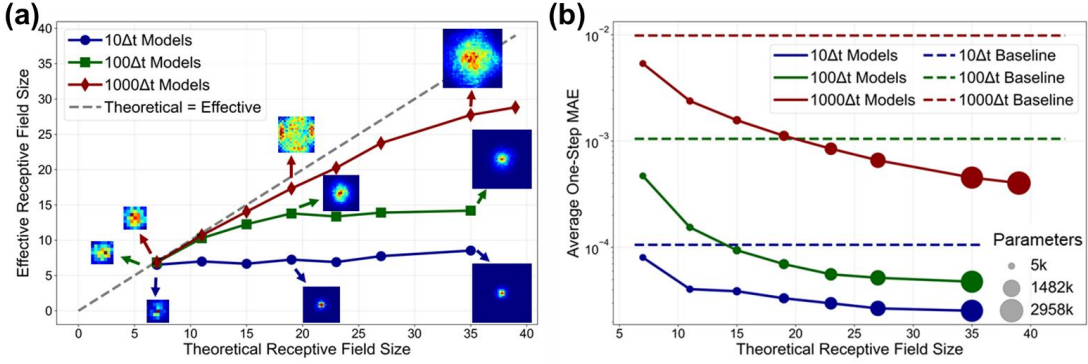

**Fig. 8.** Investigating determinants of ERFs in ideal grain growth cases. (a) Variation of $s_{ERF}$ with $s_{TRF}$ for models with different timesteps (10Δt, 100Δt, and 1000Δt), along with corresponding activation patterns. (b) Average one-step $MAE$ versus $s_{TRF}$, with circle sizes indicating model parameter counts and dashed lines representing average $IB_{total}$.

**4.2 Long-term generalization capability analysis from a reductionist perspective**

Conventional surrogate models based on LSD adopt a holistic perspective, treating entire microstructural images as wholes and attempting to approximate complete system evolution dynamics within low-dimensional latent spaces [35]. However, this holistic methodology inherently limits the model's generalization capability, making it unreliable for predicting late-stage microstructures whose global morphologies differ significantly from those within the training data's temporal range. In contrast, our results demonstrate that PSD models trained solely on short-term evolution data can reliably predict long-term microstructural evolution. Building on the local dependencies and spatiotemporal invariance of PSD models confirmed in Section 4.1, this section adopts a reductionist perspective to explain the model's remarkable long-term generalization capability and reveal its physical interpretability from a structure-energy modeling standpoint. The following analysis focuses on the ideal grain growth model, with similar analyses for non-isothermal grain growth and spinodal decomposition provided in Section S9 of the Supplementary Material.

To facilitate our analysis, we adopt the concept of "local environments" from molecular dynamics, where it describes the chemical and geometric surroundings of atoms [25,26,67]. In our context, we define the local environment of a target grid point as the multiphysics field information (e.g., phase, concentration, and temperature fields) contained within the model's ERF. As established in Section 4.1, only inputs within the ERF contribute to target grid point predictions. Therefore, from a reductionist perspective, PSD surrogate models function as regressors that predict target grid point evolution based solely on local environment information. The regressor first extracts local environment features for each grid point through multiple k×k convolutional layers in the backbone, then maps these features to next-timestep predictions via the final 1×1 convolutional layer (functionally equivalent to linear regression) and sigmoid activation function. Consistent with PF physics priors, this trained regressor, which approximates local evolution rules, applies spatially and temporally invariant operations to predict each grid point's evolution, thereby completing entire microstructural evolution predictions. From this reductionist perspective, regardless of morphological differences between early and late-stage microstructures, the reliability of the regressor's predictions for late-stage evolution depends solely on whether similar local environments were encountered in the early-stage training data. To test this hypothesis, we visualize the distribution of local environment features extracted through multiple k×k convolutions as described in Section 2.6.

We randomly sampled 100 grid points from each microstructural image in the training dataset and extracted their feature vectors for PCA dimensionality reduction. Subsequently, we performed identical feature extraction on late-stage evolution test data and projected these features into the same reduced-dimensional space. Fig. 9(a) presents the analysis results for ideal grain growth. In this plot, the central color of each scatter point indicates the predicted output value, while the edge color distinguishes between training and test set samples. Representative local environments corresponding to selected scatter points are also displayed. Different types of local environments cluster together on a single, continuous manifold within the feature space (delineated by the dashed black line), forming a highly ordered distribution along it. Specifically, intragranular environments cluster in the left region, multi-grain junction environments in the lower right corner, and grain boundary environments occupy transitional regions between them. Crucially, the local environments from the late-stage evolution test data show substantial overlap with those from training data, validating our hypothesis and explaining why the surrogate models in Fig. 2(a) can make one-step predictions for late-stage test data as reliably as for early-stage data. Fig. 9(b) further contrasts the density distributions of local environments from an early microstructure (frame 0, from the training set) and a late-stage one (frame 1000, from the test set). Although early and late-stage microstructures exhibit markedly different global morphologies, environments from late-stage data remain well within the domain spanned by training data, with only differences in density distribution. This observation indicates that microstructural evolution in surrogate models effectively represents a redistribution of a finite set of local environments within their established feature space manifold.

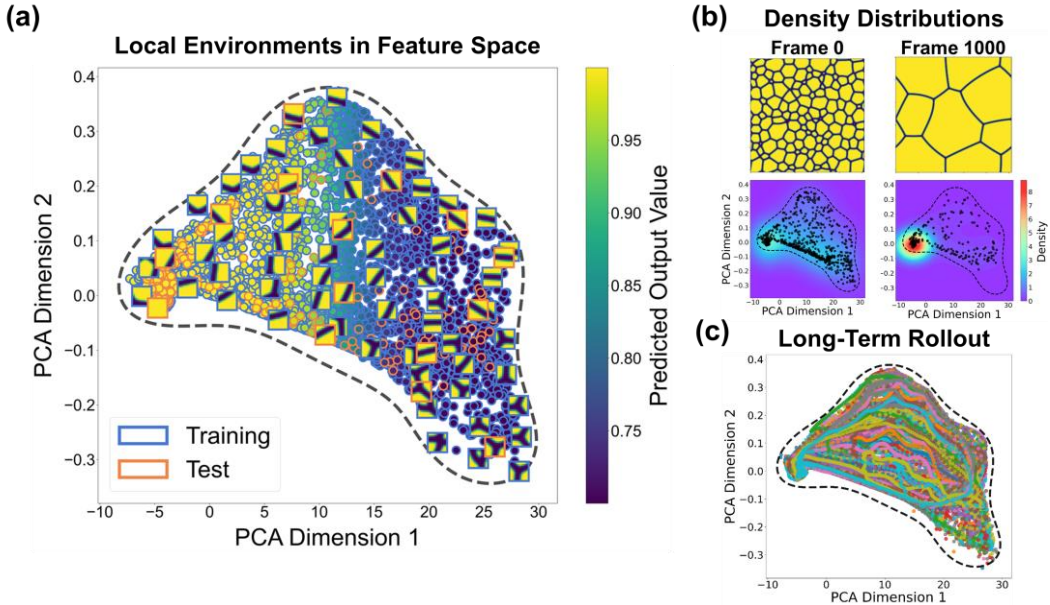

**Fig. 9.** Analysis of surrogate model's long-term generalization capability through local environment feature space. (a) Distribution of local environments from ideal grain growth training and test datasets in feature space, with representative local environments shown for selected scatter points. (b) Microstructural morphologies and corresponding local environment density distributions in feature space at frames 0 and 1000. (c) Evolution trajectories of fixed grid points in feature space during rollout inference.

The above analysis elucidates the generalization mechanism of PSD models: provided that long-term evolution does not generate new local environments beyond those in the training set, the model can make reliable one-step predictions for morphologically distinct late-stage microstructures. It is crucial to emphasize that this condition—that the set of local environments does not expand over time—holds for processes like ideal grain growth that approach steady states. However, this condition may not hold for more complex microstructural evolution processes such as dendritic solidification. Consequently, surrogate models may produce catastrophic errors when predicting local environments far from the training dataset. This challenge can be addressed through active learning [72–75] techniques, which strategically sample new data points to enhance training set representativeness. Moreover, the distance in feature space [76] described above naturally provides uncertainty estimates that can guide active learning sampling strategies, presenting a promising avenue for future exploration.

In practice, one-step generalization alone is insufficient to guarantee stable long-term predictions during autoregressive rollout. To prevent the accumulation of errors that can lead to divergence [77], the model must ensure that its predictions consistently generate local environments that remain within the manifold. To investigate rollout stability, we tracked the evolution trajectories of fixed grid points in feature space during the model's rollout inference process, as shown in Fig. 9(c). The vast majority of these trajectories remain confined within the manifold domain defined by training data, demonstrating the stability of long-term inference. It should be noted that, when a spinodal decomposition model without sigmoid boundedness constraints is used for rollout, error accumulation causes inferred local environments to gradually drift and eventually exceed the manifold domain, leading to divergence (see Supplementary Material, Fig. S6). This result highlights the critical role of embedding physical constraints into model architectures to ensure the stability of long-term rollout inference.

Potential functions are fundamental to molecular dynamics, governing atomic forces and thereby controlling system evolution. ML potentials approximate such energy functions to enable large-scale simulations [25–28]. Our analysis has revealed that microstructural evolution surrogate models learn spatiotemporally invariant local evolution rules. Here, we further propose that these learned rules function as implicit free energy landscapes that govern system evolution, providing a profound physical interpretation of the models. To visualize this landscape, we explored the relationship between local environments and their corresponding free energies within the feature space. The energy of each local environment was computed by summing free energy contributions from all constituent grid points, with individual grid point energies calculated using Eq. (S2) (Supplementary Material) for ideal grain growth.

Local environment energies are visualized in feature space, as shown in Fig. 10(a). Local energy exhibits gradual variations along the first PCA principal component, reflecting the physical law of progressively increasing interfacial energy from bulk grains to grain boundaries to grain junctions. By fitting the data points in Fig. 10(a), we obtained the "free energy surface of local environments," enabling understanding of microstructural evolution from a perspective analogous to molecular conformational transitions on potential energy surfaces in molecular dynamics simulations. As shown in Fig. 10(b), typical topological events in grain growth manifest as evolution trajectories along the free energy surface, accompanied by energy fluctuations. A grain annihilation event, for instance, appears as a transition from a low-energy grain-interior state, through a high-energy small-grain state, to decomposition into multiple triple junctions, before finally settling into a lower-energy grain boundary. A neighbor-switching event involves the shortening of a grain boundary, formation of a high-energy, unstable quadruple junction, and rapid decomposition into newly oriented grain boundaries. Combined with Fig. 9(b), the reduction in total system energy during evolution can be understood as temporal shifts in local structure probability distributions, with increasing proportions of low-energy bulk structures and decreasing proportions of high-energy interface structures. This energy perspective reveals that surrogate models implicitly learn to construct free energy landscapes that govern microstructural evolution, establishing a data-driven framework that captures underlying thermodynamic driving forces without explicit physical equations.

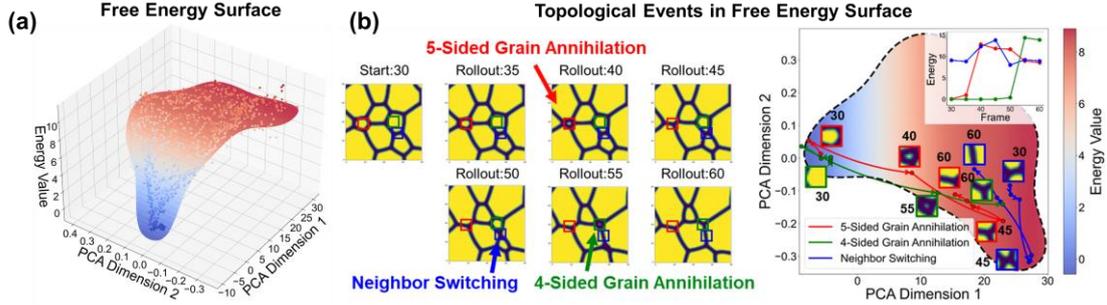

**Fig. 10.** Physical interpretation of ideal grain growth surrogate model through free energy landscapes of local environments. (a) The free energy surface is obtained by fitting local environment energies into the feature space. (b) Evolution trajectories of three typical topological events on the free energy surface, with corresponding energy variation curves illustrated in the upper right inset.

**4.3 Computational efficiency analysis**

This section evaluates the computational efficiency of the proposed ML surrogate models. The analysis considers two critical aspects: first, the upfront computational cost associated with data generation and model training, which determines the overall practicality of the methodology; and second, the computational speedup achieved during the inference stage, which quantifies the acceleration benefits. To ensure fair comparison, all computations for both PF simulations and ML surrogate models were performed on an NVIDIA A800 80GB GPU.

Previous studies on ML acceleration of PF simulations typically required tens to hundreds of microstructural evolution trajectories for training [37,39,51–54], with data generation and model training often consuming several days. In contrast, this study demonstrates that models trained on data from a single 256×256 PF simulation can achieve excellent predictive performance (as shown in Fig. 2 and Fig. 5), with the entire workflow completing within hours, just as summarized in Table 1 for various surrogate models. As discussed in Section 4.2, this efficiency arises because data collected from a single microstructural evolution trajectory provides an extensive collection of local environment-to-evolution value mappings, which adequately spans the range of possible local environments and enables robust model training. This finding suggests that data requirements for training effective ML surrogates are substantially lower than previously assumed, significantly enhancing the practical applicability of data-driven approaches.

**Table 1**

Computational costs for data collection and training of the surrogate models. Data collection time encompasses both PF simulation runtime and data storage operations.

| Model | Data collection time (s) | Model training time (h) |
|---|---|---|
| Ideal grain growth | 972 | 3.8 |
| Non-isothermal grain growth | 565 | 4.4 |
| Spinodal decomposition (phase separation) | 179 | 3 |
| Spinodal decomposition (early coarsening) | 31 | 0.6 |
| Spinodal decomposition (late coarsening) | 63 | 1.5 |

For grain growth models, we measured the computational time required for both surrogate models and PF simulations to complete $100\Delta t$ evolution. For spinodal decomposition models, we compared the computational time of three surrogate models with different timestep sizes against PF simulations for $500\Delta t$ of evolution. Fig. 11(a) and (b) present average computation times based on 500 independent runs across varying simulation system sizes. Nearly all methods exhibit linear scaling between computational time and system size, consistent with the local nature of the calculations in both PF equations and CNN surrogate models. The PF model for spinodal decomposition shows slight super-linear scaling at smaller system sizes due to the underutilization of the GPU, before transitioning to the expected linear trend. The ideal grain growth surrogate model achieves approximately 24× acceleration, while the non-isothermal grain growth surrogate model achieves 9× acceleration. However, for the computationally simpler spinodal decomposition problem, which involves only concentration field evolution, acceleration benefits are more modest: only the $500\Delta t$ surrogate model achieves 4× acceleration compared to PF simulations, the $100\Delta t$ model matches PF simulations' speed, and the $10\Delta t$ model is actually slower than direct PF simulations. These results indicate that the acceleration benefits of ML models primarily stem from their ability to permit larger timestep spans. However, determining optimal timestep size remains an open challenge. Excessively large timesteps fail to accurately capture rapid or dramatic evolution processes. Conversely, excessively small timesteps result in slower computational speeds. Additionally, while small timesteps may appear to offer lower $MAE$, they impose more stringent fitting requirements relative to smaller baselines, as shown in Fig. 8(b). Furthermore, small timesteps necessitate more iterations during rollout inference, potentially leading to more severe error accumulation [78].

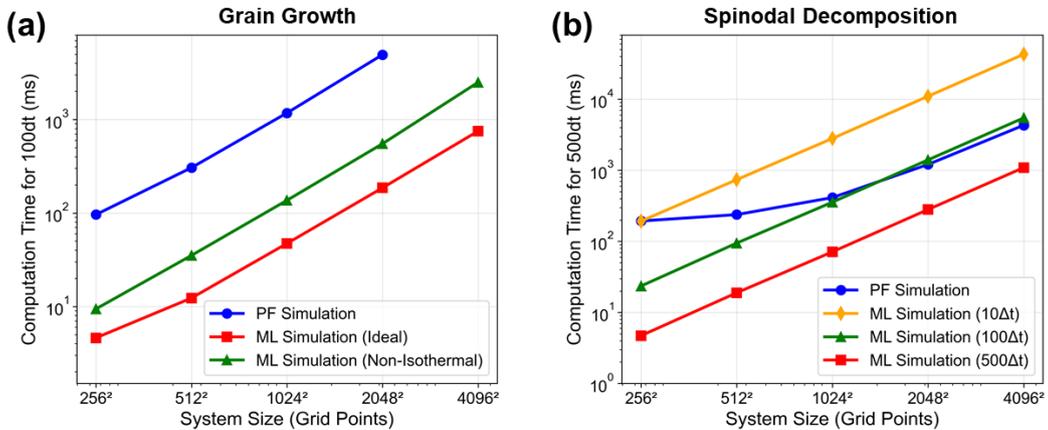

**Fig. 11.** Computational time comparison between PF models and surrogate models across different system sizes for (a) grain growth and (b) spinodal decomposition.

It should be emphasized that the primary objective of this work is to overcome the challenge of extensive data requirements and to elucidate the long-term prediction capabilities of surrogate models, rather than to achieve state-of-the-art acceleration performance. Performance was thus constrained by our choice of a parameter-intensive ResNet architecture without extensive hyperparameter optimization. Future work should focus on enhancing computational efficiency through more lightweight CNN architectures, such as those using depthwise separable convolutions [79,80]. Moreover, a fundamental limitation of CNN-based surrogate models is their requirement to perform computations on all grid points, whereas finite difference phase-field solvers can implement adaptive algorithms that compute only at interfacial regions. Therefore, exploring alternative model architectures such as graph neural networks (GNNs) [81] with adaptive mesh refinement represents another promising direction for future research [36]. Systematic benchmarking of various models is also necessary to establish state-of-the-art standards and to clarify critical trade-offs between speed and predictive accuracy.

## 5. Conclusion

In this work, we have developed and validated a data-driven framework based on PSD surrogate models to accelerate PF simulations of microstructural evolution. By demonstrating how CNN architectures with appropriate inductive biases align with the fundamental physical priors of PF simulations—namely, local dependencies and spatiotemporal translation invariance—we have shown that these models can overcome fundamental challenges of previous data-driven approaches. Our findings provide not only a practical pathway for significantly reducing the training data requirements of surrogate modeling but also offer a clear mechanistic understanding of their long-term generalization capabilities. The principal conclusions drawn from this study are as follows:

1. CNN-based surrogate models can be trained effectively using data from a single, small-scale simulation trajectory. These models achieve reliable, long-term predictions for microstructural evolution in both grain growth and spinodal decomposition, extending far beyond the temporal range of training data. Furthermore, surrogate models exhibit seamless spatial scalability, allowing those trained on small systems to accurately predict the evolution of significantly larger systems without retraining.
2. Through ERF analysis, we have quantitatively verified that surrogate models learn to make predictions based on localized information, mirroring the local nature of the PF equations. For non-isothermal grain growth, surrogate models exhibit adaptive behavior by automatically expanding their effective receptive fields in high-temperature regions where rapid evolution occurs, confirming successful capture of temperature-dependent dynamics.
3. Systematic experiments reveal that the size of ERF is determined by timestep rather than network hyperparameters, saturating at an upper limit dictated by the physical laws embedded in training data. The TRF of the network must be sufficiently large to encompass the physically determined ERF. Once the TRF accommodates the ERF, further increases in network depth and complexity yield only marginal improvements in accuracy while substantially increasing computational costs.
4. From a reductionist perspective, the remarkable long-term generalization capability of surrogate models does not stem from learning the dynamics of entire microstructures. Instead, models learn spatiotemporally translation-invariant rules that map a grid point's local environment to their future states. We have shown that the set of local environments encountered during late-stage evolution is largely contained within the distribution of environments from early-stage training data. Consequently, microstructural evolution effectively represents a redistribution of these learned local environments, enabling accurate long-term prediction.
5. The stability of long-term autoregressive inference critically depends on embedding physical priors into model architectures. Implementation of bounded output constraints was essential for preventing error accumulation and prediction divergence, as it ensures that inferred local environments remain within the manifold domain established during training.

## CrediT authorship contribution statement

**Zishuo Lan:** Writing – original draft, Investigation, Methodology, Software, Visualization, Conceptualization. **Qionghuan Zeng:** Software, Formal analysis. **Weilong Ma:** Software, Formal analysis. **Xiangju Liang:** Formal analysis. **Yue Li:** Formal analysis. **Yu Chen:** Formal analysis. **Yiming Chen:** Formal analysis. **Xiaobing Hu:** Formal analysis. **Junjie Li:** Writing – review & editing, Supervision, Methodology, Formal analysis, Funding acquisition, Conceptualization. **Lei Wang:** Formal analysis. **Jing Zhang:** Formal analysis. **Zhijun Wang:** Formal analysis. **Jincheng Wang:** Supervision, Formal analysis, Funding acquisition, Conceptualization.

## Declaration of Generative AI and AI-assisted technologies in the writing process

During the preparation of this work, the authors used ChatGPT to polish the language. After using this tool/service, the authors reviewed and edited the content as needed and took full responsibility for the publication's content.

## Declaration of competing interests

The authors declare that they have no known competing financial interests or personal relationships that could have appeared to influence the work reported in this paper.

## Acknowledgments

This work was supported by the National Natural Science Foundation of China (Grants No. 52471017), Advanced Materials-National Science and Technology Major Project (Grant No. 2025ZD0618501).

## Data availability

The data and codes are uploaded to the author's GitHub repository: https://github.com/Lan-zs/ML_acceleration_PF.